\documentclass{article}
\usepackage{spconf,amsmath,graphicx,hyperref}
\usepackage{booktabs} 
\usepackage{url} 
\usepackage{multirow}

\usepackage{array}
\usepackage{pifont}
\newcommand{\cmark}{\ding{51}}
\newcommand{\xmark}{\ding{55}}
\newcolumntype{M}{>{\centering\arraybackslash}m{1.6em}}

\newcommand{\bestaudio}[1]{\textbf{#1}}   
\newcommand{\overallbest}[1]{\underline{\textbf{#1}}} 

\usepackage{background}
\SetBgAngle{0}
\SetBgScale{1}
\SetBgColor{black}
\SetBgOpacity{1}
\SetBgPosition{.51\paperwidth,-1.08\textheight}
\SetBgContents{
{\parbox{\paperwidth}{%
\footnotesize \copyright 2025 IEEE.  Personal use of this material is permitted. 
Permission from IEEE must be obtained for all other uses, in any current or future media,\\
including reprinting/republishing this material for advertising or 
promotional purposes, creating new collective works, for resale or redistribution\\
to servers or lists, or reuse of any copyrighted component of this work in other works.
\mbox{}\\
     \mbox{}\\}
     }%
}

\title{SpeechCT-CLIP: Distilling Text-Image Knowledge to Speech for Voice-Native Multimodal CT Analysis}

\name{%
\begin{tabular}{@{}c@{}}
 Lukas Buess$^{1,*,\dagger}$, Jan Geier$^{1,*}$, David Bani-Harouni$^{2}$,
Chantal Pellegrini$^{2}$, Matthias Keicher$^{2}$, \\
Paula Andrea Perez-Toro$^{1}$, Nassir Navab$^{2}$,
Andreas Maier$^{1}$, Tomas Arias-Vergara$^{1}$\thanks{The authors gratefully acknowledge the scientific support and HPC resources provided by the Erlangen National High Performance Computing Center (NHR@FAU) of the Friedrich\mbox{-}Alexander\mbox{-}Universit\"at Erlangen\mbox{-}N\"urnberg (FAU). The hardware is funded by the German Research Foundation (DFG). This work was partially funded via the EVUK programme (``Next-generation AI for Integrated Diagnostics'') of the Free State of Bavaria and the Deutsche Forschungsgemeinschaft (DFG).}
\end{tabular}
}

\address{$^{1}$ Pattern Recognition Lab, Friedrich\mbox{-}Alexander\mbox{-}Universit\"at Erlangen\mbox{-}N\"urnberg, Erlangen, Germany \\
         $^{2}$ Computer Aided Medical Procedures, Technical University of Munich, Munich, Germany \\
         $^{*}$ These authors contributed equally to this work. \\
         $^{\dagger}$ Corresponding author: Lukas.Buess@fau.de}

\begin{document}
%
\maketitle
\begin{abstract}
Spoken communication plays a central role in clinical workflows. In radiology, for example, most reports are created through dictation. Yet, nearly all medical AI systems rely exclusively on written text. In this work, we address this gap by exploring the feasibility of learning visual-language representations directly from spoken radiology reports. Specifically, we synthesize a large-scale dataset (Speech-RATE) of spoken radiology reports and train SpeechCT-CLIP, a contrastive model that aligns speech and 3D CT volumes in a shared representation space. While naïve speech-based models underperform compared to text-trained counterparts, we show that knowledge distillation from a pretrained text-image CLIP model effectively transfers semantic alignment capabilities from text to speech, substantially narrowing this gap. Experiments demonstrate improved zero-shot classification F1 from 0.623 to 0.705, recovering 88\% of the performance difference, and strong retrieval results without requiring text at inference. These findings highlight speech as a practical alternative to text in multimodal pretraining and open the door to voice-driven diagnostic support tools in clinical practice.
\end{abstract}

\section{Introduction}
\label{sec:intro}
Recent advances in multimodal learning have transformed medical image analysis. By pairing visual data with unstructured clinical reports, contrastive pretraining has yielded robust foundation models capable of retrieval, abnormality detection, and even report generation \cite{hamamci2024developing,hamamci2024ct2rep, buess2025large}. The release of CT-RATE \cite{hamamci2024developing,hamamci2024generatect}, the first large-scale dataset linking 3D CT volumes to radiology reports, enabled CT-CLIP, a vision-language model (VLM) that demonstrated strong zero-shot performance across a wide range of tasks. These successes underscore the central role of natural language in shaping visual representations for radiology.

However, radiologists rarely directly write reports, instead dictation remains the dominant reporting modality in everyday practice: radiologists scroll through CT volumes and describe their findings verbally, with automatic speech recognition (ASR) systems transcribing the audio into text for the electronic health record (EHR). Despite being widespread, ASR is not perfect and can lead to clinically significant errors \cite{schmidt2024generative}. Such errors not only burden clinicians with correction work but also pose risks to patient safety. These challenges motivate a different question: instead of relying on transcription, could we build models that understand radiology reports directly from speech?

In computer vision, CLIP has been extended to audio: AudioCLIP \cite{guzhov2022audioclip} aligns audio with images and text via contrastive learning, and Wav2CLIP \cite{wu2022wav2clip} distills CLIP’s embedding space into an audio encoder. Large-scale audio-language resources such as Auto-ACD \cite{sun2024auto} further enable robust audio-text representation learning. In biomedicine, domain-specific vision-language pretraining with BiomedCLIP \cite{zhang2024multimodal} demonstrated strong cross-modal transfer across retrieval and classification and ECG-Text Pretraining (ETP) \cite{liu2024etp} extended text-paired contrastive pretraining to physiological signals. Complementary to these, acoustic-driven clinical reporting has been explored by generating pathological speech reports from audio with LLMs \cite{arias2025acoustic}. Despite this progress, existing medical contrastive models rely on text supervision; alignment of spoken radiology reports with medical images remains unaddressed.

\begin{figure*}[ht]
    \centering
    \includegraphics[width=0.8\textwidth]{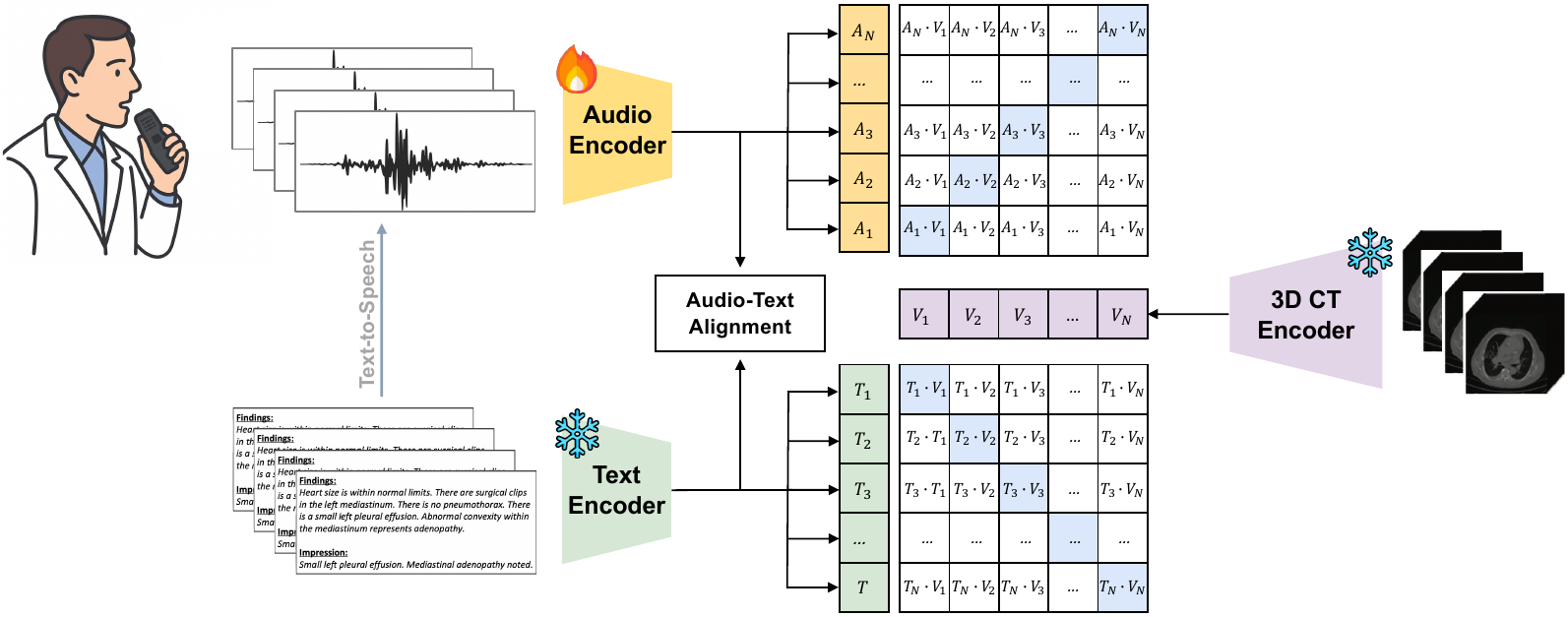}
    \caption{Overview of SpeechCT-CLIP. A frozen, pretrained CT and text encoder supervise the audio encoder. Spoken reports yield audio embeddings \(\{A_i\}\), aligned with CT features \(\{V_j\}\) via contrastive learning and distilled from text embeddings \(\{T_i\}\).}
    \label{fig:overview}
\end{figure*}

To bridge this gap we introduce SpeechCT-CLIP, the first foundation model that aligns spoken radiology reports with 3D CT volumes. To build this system, we synthesize a large-scale dataset (Speech-RATE) of spoken radiology reports from CT-RATE using diverse synthetic voices, and train a speech encoder alongside a CT encoder in a contrastive setup. While naïve training reveals a significant performance drop compared to CT-CLIP (text-based), we propose a simple yet effective solution: knowledge distillation from the stronger text encoder. By transferring semantic knowledge from text embeddings to audio embeddings during training, we substantially narrow the gap while enabling inference directly from speech. Our contributions are as follows:

\begin{itemize}
    \item We introduce Speech-RATE, the first large-scale dataset of spoken radiology reports paired with CT volumes from CT-RATE \cite{hamamci2024developing}, totaling 50,188 synthetic spoken reports with diverse voices and dictation styles.
    \item We establish baselines for speech-CT contrastive learning and direct spoken report classification, providing the first benchmarks of their kind in medical imaging.
    \item We introduce text-guided distillation that transfers knowledge from a text-based CLIP teacher to a speech-based student, enabling robust speech-only inference.
\end{itemize}
Our results highlight both the challenges and opportunities of replacing text with speech in medical multimodal pretraining, opening the door to robust, voice-native medical AI systems.

\section{Method}
We propose SpeechCT-CLIP, a contrastive vision-speech model that aligns spoken radiology reports with 3D~CT volumes in a shared representation space. A pretrained image encoder is aligned with a trainable speech encoder using contrastive and distillation objectives (see Figure~\ref{fig:overview}). We first describe the construction of the Speech-RATE dataset, and then detail the training method.

\subsection{Speech-RATE: Synthetic Spoken Report Dataset}
\label{sec:method:dataset}
We construct Speech-RATE by extending CT-RATE \cite{hamamci2024developing}, which consists of 50,188 3D CT volumes, each paired with a written radiology report and labeled across 18 abnormality labels. To simulate radiologist dictation, each report’s findings section is converted to audio using the Kokoro TTS engine\footnote{\scriptsize \url{https://huggingface.co/hexgrad/Kokoro-82M}}.

Speech is synthesized with eight distinct voices (4 female, 4 male), with small random variations in speaking rate to mimic realistic dictation. All clips are resampled to 24\,kHz. The resulting dataset comprises 50,188 spoken findings sections totaling 1,197\,hours of audio with an average length of 86\,s (see Table~\ref{tab:dataset}). To support future research, Speech-RATE is released publicly on Hugging Face\footnote{\scriptsize Dataset will be released upon paper acceptance.}.

\begin{table}[h]
  \centering
  \vspace{-6pt}
  \caption{Statistics of the Speech-RATE dataset.}
  \label{tab:dataset}
  \begin{tabular}{l@{\hskip 4em}c}
    \toprule
    \textbf{Property} & \textbf{Value} \\
    \midrule
    Spoken findings sections & 50,188 \\
    Total duration & 1,197\,h \\
    Avg. length & 86\,s \\
    Language & English \\
    Voices & 8 (4F / 4M) \\
    TTS engine & Kokoro \\
    Sampling rate & 24\,kHz \\
    Augmentations & Speed \\
    \bottomrule
  \end{tabular}
\end{table}

\subsection{Contrastive Pretraining}
\label{sec:method:contrastive}
Following the CT-CLIP framework \cite{hamamci2024developing,radford2021learning}, SpeechCT-CLIP learns a joint representation space for CT volumes and spoken reports. Let $g(\cdot)$ denote the CT-ViT encoder \cite{hamamci2024generatect} used in CT-CLIP \cite{hamamci2024developing}, which we adopt as the vision backbone, and $f_{\theta}(\cdot)$ the trainable speech encoder. For each CT volume $x^{\text{ct}}_i$ and its paired audio waveform $x^{\text{a}}_i$, we compute
\begin{equation}
v_i = g(x^{\text{ct}}_i), \quad a_i = f_{\theta}(x^{\text{a}}_i),
\end{equation}
where $v_i$ and $a_i$ represent the corresponding embeddings in the shared latent space.

Since the audio encoder only accepts a limited input length and many spoken reports are long, we use a sliding-window strategy. Each waveform is divided into windows of length $L=30$\,s, with an overlap of $O=2$\,s. These are encoded separately by $f_{\theta}$ and then averaged to form the final embedding $a_i$.

To align audio and CT embeddings, we employ a contrastive objective. For a minibatch of $N$ paired examples, the loss for matching audio to CT is
\begin{equation}
\mathcal{L}_{\text{con}} = -\frac{1}{N}\sum_{i=1}^N 
\log \frac{\exp(\cos(a_i,v_i))}{\sum_{j=1}^N \exp(\cos(a_i,v_j))},
\end{equation}
where $\cos(\cdot,\cdot)$ denotes cosine similarity. In practice, we use the symmetric version of this loss, summing audio$\to$CT and CT$\to$audio terms.

\subsection{Knowledge Distillation}
Direct speech-CT alignment described in \ref{sec:method:contrastive} lags behind text-CT alignment. To reduce this gap, we use the pretrained CT-CLIP text encoder $h(\cdot)$ as a teacher, while keeping both CT and text encoders frozen. For each report $x^{\text{t}}_i$, we obtain the teacher embedding
\begin{equation}
t_i = h(x^{\text{t}}_i).
\end{equation}
The speech embedding $a_i$ is then encouraged to match $t_i$ using cosine similarity:
\begin{equation}
\mathcal{L}_{\text{distill}} = 1 - \text{cos}(a_i, t_i).
\end{equation} 

The final training objective combines contrastive alignment with distillation:
\begin{equation}
\mathcal{L} = \mathcal{L}_{\text{con}} + \lambda \,\mathcal{L}_{\text{distill}},
\end{equation}
with $\lambda$ balancing the two terms.

\subsection{Inference}
At inference, only the CT encoder $g(\cdot)$ and the speech encoder $f_{\theta}(\cdot)$ are active, enabling voice-native applications such as case retrieval, abnormality classification, or integration into multimodal assistants without intermediate text transcription.

\section{Experimental Setup}

\subsection{Data Setup}
We use Speech-RATE (Section~\ref{sec:method:dataset}), our synthetic dataset of spoken radiology reports paired with CT volumes from CT-RATE \cite{hamamci2024developing}. We use the official train/test splits and the lightweight RadGenome-ChestCT version \cite{zhang2024radgenome} for efficiency. To test robustness to speaker variation and unseen images, we evaluate on RAD-ChestCT \cite{draelos2021machine}, containing scans from another hospital. Following prior work \cite{hamamci2024developing}, arterial and coronary artery wall calcification are merged and mosaic attenuation removed, resulting in 16 labels. For each item, speech is synthesized using a random voice from 8 unseen speakers.

\subsection{Evaluation}
We first evaluate zero-shot abnormality classification on the 18 CT-RATE labels using AUROC, F1, accuracy, and precision (averaged across labels). In addition, we assess cross-modal case retrieval with Recall@$\{5,10,50,100\}$.

\subsection{Implementation Details}
Models are trained on a single NVIDIA A100 (80\,GB) GPU using the Adam optimizer with a learning rate of $1.25 \times 10^{-6}$ and a batch size of 8 for 100{,}000 steps.

\section{Results and Discussion}
\begin{figure}[h]
    \centering
    \includegraphics[width=\columnwidth]{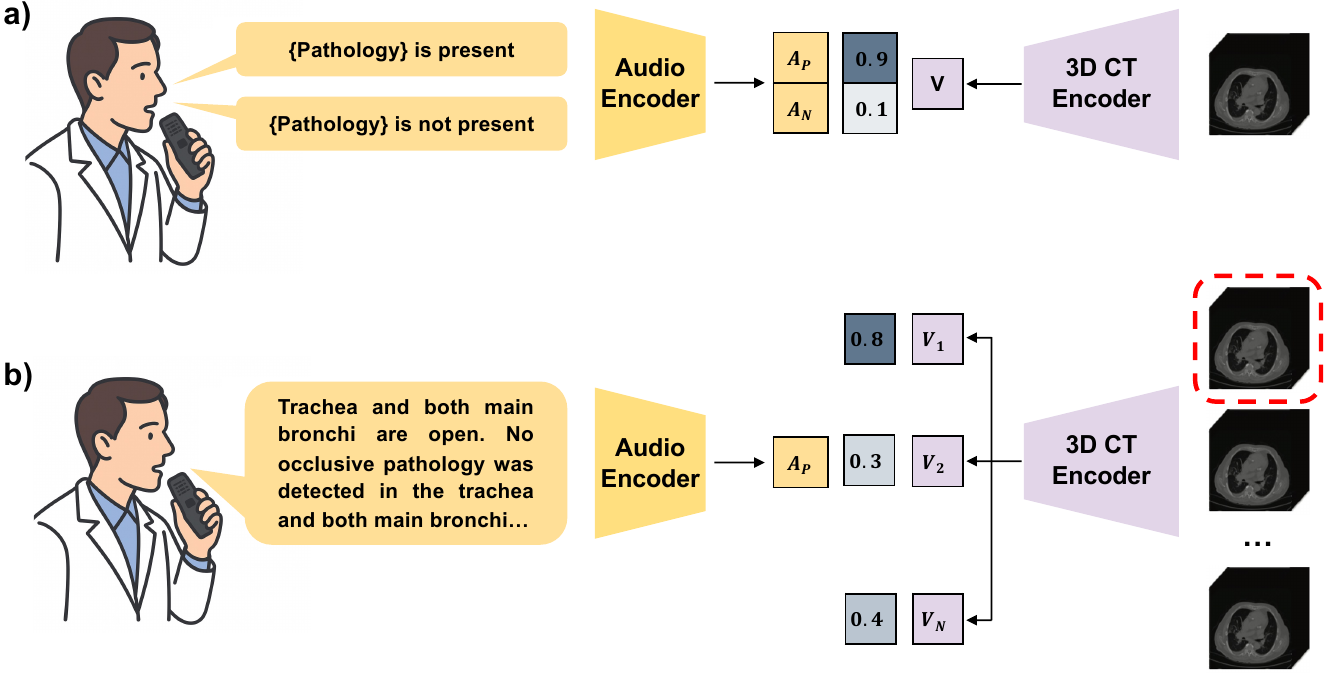}
    \caption{(a) Zero-shot classification: spoken pathology queries are matched with CT features. (b) Case retrieval: spoken reports are used to retrieve corresponding CT volumes.}
    \label{fig:downstream_tasks}
\end{figure}

We evaluate SpeechCT-CLIP on zero-shot classification and cross-modal retrieval (Figure~\ref{fig:downstream_tasks}). As baselines, we report Random predictions, CT-Net \cite{hamamci2024developing} as a supervised vision-only model, and CT-CLIP \cite{hamamci2024developing} as a text-based contrastive model. For speech, we compare a naïve variant, SpeechCT-CLIP$_{\text{nKD}}$ (no knowledge distillation), trained only with contrastive loss, against our full SpeechCT-CLIP, which adds text-guided distillation. This progression isolates the roles of vision-only training, text supervision, speech alignment, and distillation.

\subsection{Speech Encoder Selection}
To select a speech encoder for SpeechCT-CLIP, we evaluated wav2vec \cite{baevski2020wav2vec}, HuBERT \cite{hsu2021hubert}, and Whisper \cite{radford2023robust} on classifying the 18 CT-RATE abnormality labels directly from the spoken reports (Table~\ref{tab:speech_encoders}), without using images. Whisper achieved the best linear probing results and improved further when finetuning its top layers. This likely reflects its pretraining on large-scale ASR, which emphasizes semantic content over acoustic detail, making it better suited for reporting. We therefore use Whisper (base) as backbone for subsequent experiments.

\begin{table}[h]
  \centering
  \vspace{-6pt}
  \caption{Comparison of pretrained speech encoders on spoken report classification. Results are weighted F1, precision (Prec.), and recall (Rec.) under linear probing and finetuning.}
  \label{tab:speech_encoders}
  \setlength{\tabcolsep}{6pt}
  \renewcommand{\arraystretch}{1.1}
  \begin{tabular*}{\columnwidth}{@{\extracolsep{\fill}} l c c c @{}}
    \toprule
    \textbf{Model} & \textbf{F1} & \textbf{Prec.} & \textbf{Rec.} \\
    \midrule
    \multicolumn{4}{c}{\textit{Linear Probing}} \\
    \midrule
    wav2vec \cite{baevski2020wav2vec} & 0.51 & 0.64 & 0.45 \\
    HuBERT \cite{hsu2021hubert} & 0.65 & \bestaudio{0.79} & 0.60 \\
    Whisper \cite{radford2023robust} & \bestaudio{0.67} & \bestaudio{0.79} & \bestaudio{0.62} \\
    \midrule
    \multicolumn{4}{c}{\textit{Finetuning}} \\
    \midrule
    Whisper (top 2 layers) & \bestaudio{0.76} & \bestaudio{0.84} & \bestaudio{0.71} \\
    Whisper (top 3 layers) & 0.75 & 0.82 & \bestaudio{0.71} \\
    \bottomrule
  \end{tabular*}
\end{table}

\subsection{Zero-Shot Classification}
We follow the evaluation protocol of CT-CLIP \cite{hamamci2024developing} and assess zero-shot classification across the 18 CT-RATE labels, reporting AUROC, F1, accuracy, and precision (Table~\ref{tab:zero_shot_modality}).

\begin{table}[h]
  \centering
  \vspace{-6pt}
  \caption{Zero-shot multi-label classification on CT-RATE (internal) and RAD-ChestCT (external). Metrics: AUC (AUROC), F1, Acc. (accuracy), Prec. (precision). Modalities at inference time: CT:\includegraphics[width=0.3cm]{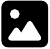}, Text:\includegraphics[width=0.25cm]{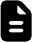}, Audio:\includegraphics[width=0.35cm]{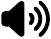}. \bestaudio{Bold} = best within audio modality; \overallbest{underlined} = overall best.}
  \label{tab:zero_shot_modality}
  \setlength{\tabcolsep}{3pt}
  \renewcommand{\arraystretch}{1.1}
  \resizebox{\columnwidth}{!}{%
    \begin{tabular}{l c c c c c c c}
      \toprule
      \multirow{2}{*}{\textbf{Model}} & \multicolumn{3}{c}{\textbf{Inference}} & \multicolumn{4}{c}{\textbf{Metrics}} \\
      \cmidrule(lr){2-4} \cmidrule(lr){5-8}
       & \includegraphics[width=0.3cm]{image.pdf}
       & \includegraphics[width=0.25cm]{text.pdf}
       & \includegraphics[width=0.35cm]{audio.pdf}
       & \textbf{AUC} & \textbf{F1} & \textbf{Acc.} & \textbf{Prec.} \\
      \midrule
      \multicolumn{8}{c}{\textit{CT-RATE (internal validation)}} \\
      \midrule
      Random\textsuperscript{*} & \xmark & \xmark & \xmark & 0.505 & 0.570 & 0.502 & 0.180 \\
      CT-Net\textsuperscript{*} \cite{hamamci2024developing} & \cmark & \xmark & \xmark & 0.629 & 0.657 & 0.617 & 0.263 \\
      CT-CLIP \cite{hamamci2024developing} & \cmark & \cmark & \xmark & \overallbest{0.734} & \overallbest{0.718} & \overallbest{0.681} & \overallbest{0.326} \\
      SpeechCT-CLIP$_{\text{nKD}}$ & \cmark & \xmark & \cmark & 0.610 & 0.623 & 0.574 & 0.248 \\
      SpeechCT-CLIP & \cmark & \xmark & \cmark & \bestaudio{0.708} & \bestaudio{0.705} & \bestaudio{0.666} & \bestaudio{0.314} \\
      \midrule
      \multicolumn{8}{c}{\textit{RAD-ChestCT (external validation)}} \\
      \midrule
      Random\textsuperscript{*} & \xmark & \xmark & \xmark & 0.496 & 0.555 & 0.500 & 0.265 \\
      CT-Net\textsuperscript{*} \cite{hamamci2024developing} & \cmark & \xmark & \xmark & 0.544 & 0.564 & 0.517 & 0.282 \\
      CT-CLIP \cite{hamamci2024developing} & \cmark & \cmark & \xmark & \overallbest{0.643} & \overallbest{0.660} & \overallbest{0.615} & \overallbest{0.343} \\
      SpeechCT-CLIP$_{\text{nKD}}$ & \cmark & \xmark & \cmark & 0.552 & 0.596 & 0.544 & 0.297 \\
      SpeechCT-CLIP & \cmark & \xmark & \cmark & \bestaudio{0.603} & \bestaudio{0.623} & \bestaudio{0.575} & \bestaudio{0.319} \\
      \bottomrule
      \multicolumn{8}{l}{\textsuperscript{*} Metrics cited from the original publication \cite{hamamci2024developing}.}
    \end{tabular}%
  }
\end{table}

On internal validation, CT-CLIP provides the strongest baseline with an F1 of 0.718, while SpeechCT-CLIP$_{\text{nKD}}$ drops to 0.623, reflecting the difficulty of learning directly from speech. Our proposed SpeechCT-CLIP substantially narrows this gap, improving F1 to 0.705 and AUROC to 0.708, thereby recovering 88\% of the performance difference between text- and speech-based models.

To assess generalization, we evaluate SpeechCT-CLIP on RAD-ChestCT. It improves over SpeechCT-CLIP$_{\text{nKD}}$ (F1 0.623 vs.\ 0.596) and approaches CT-CLIP (F1 0.660), demonstrating robustness to new domains and speakers.

\subsection{Case Retrieval}
We further evaluate cross-modal retrieval, where spoken reports are used to retrieve similar CT volumes (Table~\ref{tab:retrieval_modality}). CT-CLIP achieves the highest recall (R@10 = 0.085, R@100 = 0.430), while SpeechCT-CLIP$_{\text{nKD}}$ underperforms due to weaker speech embeddings. SpeechCT-CLIP improves retrieval substantially (R@10 = 0.077, R@100 = 0.377), closing much of the gap to text-based performance and confirming the benefit of distillation for speech-image alignment.

\begin{table}[h]
  \centering
  \vspace{-6pt}
  \caption{Cross-modal retrieval on CT-RATE. Modalities at inference time: CT:\includegraphics[width=0.3cm]{image.pdf}, Text:\includegraphics[width=0.25cm]{text.pdf}, Audio:\includegraphics[width=0.35cm]{audio.pdf}. \bestaudio{Bold} = best within audio modality; \overallbest{underlined} = overall best.}
  \label{tab:retrieval_modality}
  \setlength{\tabcolsep}{3pt}
  \renewcommand{\arraystretch}{1.1}
  \resizebox{\columnwidth}{!}{%
    \begin{tabular}{l c c c c c c c}
      \toprule
      \multirow{2}{*}{\textbf{Model}} & \multicolumn{3}{c}{\textbf{Inference}} & \multicolumn{4}{c}{\textbf{Recall}} \\
      \cmidrule(lr){2-4} \cmidrule(lr){5-8}
       & \includegraphics[width=0.3cm]{image.pdf}
       & \includegraphics[width=0.25cm]{text.pdf}
       & \includegraphics[width=0.35cm]{audio.pdf}
       & \textbf{@5} & \textbf{@10} & \textbf{@50} & \textbf{@100} \\
      \midrule
      Random & \xmark & \xmark & \xmark & 0.003 & 0.005 & 0.036 & 0.056 \\
      CT-CLIP \cite{hamamci2024developing} & \cmark & \cmark & \xmark & \overallbest{0.048} & \overallbest{0.085} & \overallbest{0.281} & \overallbest{0.430} \\
      SpeechCT-CLIP$_{\text{nKD}}$ & \cmark & \xmark & \cmark & 0.026 & 0.049 & 0.180 & 0.291 \\
      SpeechCT-CLIP & \cmark & \xmark & \cmark & \bestaudio{0.042} & \bestaudio{0.077} & \bestaudio{0.244} & \bestaudio{0.377} \\
      \bottomrule
    \end{tabular}%
  }
\end{table}

\section{Conclusion}
In this work, we proposed SpeechCT-CLIP, the first model to align spoken radiology reports with 3D CT volumes. Building on our newly created Speech-RATE dataset, we demonstrated that speech can serve as a promising alternative to text in multimodal pretraining. By distilling knowledge from a pretrained text encoder, SpeechCT-CLIP substantially narrows the gap between text- and speech-based training. This enables robust voice-native medical AI without intermediate transcription, avoiding ASR errors and preserving natural information such as radiologists’ uncertainty that would otherwise be lost in transcription. Our work opens the path towards applications beyond classification and retrieval, including speech-driven segmentation, multilingual training, and integration into interactive clinical assistants. In doing so, SpeechCT-CLIP paves the way for more natural and accessible human-AI interaction in clinical practice.

\vfill\pagebreak

\bibliographystyle{IEEEbib}
\bibliography{main}

\begin{thebibliography}{10}

\bibitem{hamamci2024developing}
Ibrahim~Ethem Hamamci, Sezgin Er, Furkan Almas, Ayse~Gulnihan Simsek, Sevval~Nil Esirgun, Irem Dogan, Muhammed~Furkan Dasdelen, Omer~Faruk Durugol, Bastian Wittmann, Tamaz Amiranashvili, et~al.,
\newblock ``Developing generalist foundation models from a multimodal dataset for 3d computed tomography,''
\newblock {\em arXiv preprint arXiv:2403.17834}, 2024.

\bibitem{hamamci2024ct2rep}
Ibrahim~Ethem Hamamci, Sezgin Er, and Bjoern Menze,
\newblock ``Ct2rep: Automated radiology report generation for 3d medical imaging,''
\newblock in {\em International Conference on Medical Image Computing and Computer-Assisted Intervention}. Springer, 2024, pp. 476--486.

\bibitem{buess2025large}
Lukas Buess, Matthias Keicher, Nassir Navab, Andreas Maier, and Soroosh Tayebi~Arasteh,
\newblock ``From large language models to multimodal ai: A scoping review on the potential of generative ai in medicine,''
\newblock {\em Biomedical Engineering Letters}, pp. 1--19, 2025.

\bibitem{hamamci2024generatect}
Ibrahim~Ethem Hamamci, Sezgin Er, Anjany Sekuboyina, Enis Simsar, Alperen Tezcan, Ayse~Gulnihan Simsek, Sevval~Nil Esirgun, Furkan Almas, Irem Do{\u{g}}an, Muhammed~Furkan Dasdelen, et~al.,
\newblock ``Generatect: Text-conditional generation of 3d chest ct volumes,''
\newblock in {\em European Conference on Computer Vision}. Springer, 2024, pp. 126--143.

\bibitem{schmidt2024generative}
Reuben~A Schmidt, Jarrel~CY Seah, Ke~Cao, Lincoln Lim, Wei Lim, and Justin Yeung,
\newblock ``Generative large language models for detection of speech recognition errors in radiology reports,''
\newblock {\em Radiology: Artificial Intelligence}, vol. 6, no. 2, pp. e230205, 2024.

\bibitem{guzhov2022audioclip}
Andrey Guzhov, Federico Raue, J{\"o}rn Hees, and Andreas Dengel,
\newblock ``Audioclip: Extending clip to image, text and audio,''
\newblock in {\em ICASSP 2022-2022 IEEE International Conference on Acoustics, Speech and Signal Processing (ICASSP)}. IEEE, 2022, pp. 976--980.

\bibitem{wu2022wav2clip}
Ho-Hsiang Wu, Prem Seetharaman, Kundan Kumar, and Juan~Pablo Bello,
\newblock ``Wav2clip: Learning robust audio representations from clip,''
\newblock in {\em ICASSP 2022-2022 IEEE International Conference on Acoustics, Speech and Signal Processing (ICASSP)}. IEEE, 2022, pp. 4563--4567.

\bibitem{sun2024auto}
Luoyi Sun, Xuenan Xu, Mengyue Wu, and Weidi Xie,
\newblock ``Auto-acd: A large-scale dataset for audio-language representation learning,''
\newblock in {\em Proceedings of the 32nd ACM International Conference on Multimedia}, 2024, pp. 5025--5034.

\bibitem{zhang2024multimodal}
Sheng Zhang, Yanbo Xu, Naoto Usuyama, Hanwen Xu, Jaspreet Bagga, Robert Tinn, Sam Preston, Rajesh Rao, Mu~Wei, Naveen Valluri, et~al.,
\newblock ``A multimodal biomedical foundation model trained from fifteen million image--text pairs,''
\newblock {\em NEJM AI}, p. AIoa2400640, 2024.

\bibitem{liu2024etp}
Che Liu, Zhongwei Wan, Sibo Cheng, Mi~Zhang, and Rossella Arcucci,
\newblock ``Etp: Learning transferable ecg representations via ecg-text pre-training,''
\newblock in {\em ICASSP 2024-2024 IEEE International Conference on Acoustics, Speech and Signal Processing (ICASSP)}. IEEE, 2024, pp. 8230--8234.

\bibitem{arias2025acoustic}
Tomas Arias-Vergara, Lukas Buess, Nastassia Vysotskaya, Soroosh~Tayebi Arasteh, Juan~Rafael Orozco-Arroyave, Maria Schuster, Elmar Noeth, Andreas Maier, and Paula~Andrea Perez-Toro,
\newblock ``Acoustic-driven generation of pathological speech reports using large language models,''
\newblock {\em Research Square}, 2025,
\newblock Preprint.

\bibitem{radford2021learning}
Alec Radford, Jong~Wook Kim, Chris Hallacy, Aditya Ramesh, Gabriel Goh, Sandhini Agarwal, Girish Sastry, Amanda Askell, Pamela Mishkin, Jack Clark, et~al.,
\newblock ``Learning transferable visual models from natural language supervision,''
\newblock in {\em International conference on machine learning}. PMLR, 2021, pp. 8748--8763.

\bibitem{zhang2024radgenome}
Xiaoman Zhang, Chaoyi Wu, Ziheng Zhao, Jiayu Lei, Ya~Zhang, Yanfeng Wang, and Weidi Xie,
\newblock ``Radgenome-chest ct: A grounded vision-language dataset for chest ct analysis,''
\newblock {\em arXiv preprint arXiv:2404.16754}, 2024.

\bibitem{draelos2021machine}
Rachel~Lea Draelos, David Dov, Maciej~A Mazurowski, Joseph~Y Lo, Ricardo Henao, Geoffrey~D Rubin, and Lawrence Carin,
\newblock ``Machine-learning-based multiple abnormality prediction with large-scale chest computed tomography volumes,''
\newblock {\em Medical image analysis}, vol. 67, pp. 101857, 2021.

\bibitem{baevski2020wav2vec}
Alexei Baevski, Yuhao Zhou, Abdelrahman Mohamed, and Michael Auli,
\newblock ``wav2vec 2.0: A framework for self-supervised learning of speech representations,''
\newblock {\em Advances in neural information processing systems}, vol. 33, pp. 12449--12460, 2020.

\bibitem{hsu2021hubert}
Wei-Ning Hsu, Benjamin Bolte, Yao-Hung~Hubert Tsai, Kushal Lakhotia, Ruslan Salakhutdinov, and Abdelrahman Mohamed,
\newblock ``Hubert: Self-supervised speech representation learning by masked prediction of hidden units,''
\newblock {\em IEEE/ACM transactions on audio, speech, and language processing}, vol. 29, pp. 3451--3460, 2021.

\bibitem{radford2023robust}
Alec Radford, Jong~Wook Kim, Tao Xu, Greg Brockman, Christine McLeavey, and Ilya Sutskever,
\newblock ``Robust speech recognition via large-scale weak supervision,''
\newblock in {\em International conference on machine learning}. PMLR, 2023, pp. 28492--28518.

\end{thebibliography}

\end{document}